\documentclass[conference]{IEEEtran}
\IEEEoverridecommandlockouts
\usepackage{cite}
\usepackage{amsmath,amssymb,amsfonts}
\usepackage{algorithmic}
\usepackage{graphicx}
\usepackage{textcomp}
\usepackage{xcolor}
\usepackage{multirow, booktabs}
\usepackage{listings,multicol}
\usepackage{xfrac}
\def\BibTeX{{\rm B\kern-.05em{\sc i\kern-.025em b}\kern-.08em
\kern-.1667em\lower.7ex\hbox{E}\kern-.125emX}}
\usepackage[utf8]{inputenc}
\usepackage[T1]{fontenc}

\begin{document}

\title{Privacy Parameter Variation using RAPPOR on a Malware Dataset}

\author{\IEEEauthorblockN{Peter Aaby,
Juan José Mata de Acuña, Richard Macfarlane and
William J Buchanan}
\IEEEauthorblockA{School of Computing,
Edinburgh Napier University, Edinburgh, UK\\
}
}

\maketitle

\begin{abstract}
Stricter data protection regulations and the poor application of privacy protection techniques have resulted in a requirement for data-driven companies to adopt new methods of analysing sensitive user data. The RAPPOR (Randomized Aggregatable Privacy-Preserving Ordinal Response) method adds parameterised noise, which must be carefully selected to maintain adequate privacy without losing analytical value. This paper applies RAPPOR privacy parameter variations against a public dataset containing a list of running Android applications data. The dataset is filtered and sampled into small (10,000); medium (100,000); and large (1,200,000) sample sizes while applying RAPPOR with $\epsilon$ = 10; 1.0; and 0.1 (respectively low; medium; high privacy guarantees). Also, in order to observe detailed variations within high to medium privacy guarantees ($\epsilon$ = 0.5 to 1.0), a second experiment is conducted by progressively adjusting the value of $\epsilon$ over the same populations. The first experiment verifies the original RAPPOR studies using $\epsilon$ = 1 with a non-existent recoverability in the small sample size, and detectable signal from medium to large sample sizes as also demonstrated in the original RAPPOR paper. Further results, using high privacy guarantees, show that the large sample size, in contrast to medium, suffers 2.75 times more in terms of recoverability when increasing privacy guarantees from $\epsilon$ = 1.0 to 0.8. Overall, the paper demonstrates that high privacy guarantees to restrict the analysis only to the most dominating strings.
\end{abstract}

\begin{IEEEkeywords}
Privacy Parameter Variation, Privacy Preservation, Big Data, RAPPOR
\end{IEEEkeywords}

\section{Introduction}\label{introduction}
Big data and the sharing of research data containing Personal Private Information (PII) such as medical data is growing, and the necessity of protecting citizens privacy follows \cite{Morran2016,ElEmam2011}. The General Data Protection Regulations (GDPR) also widens the scope of PII and increases the responsibility of data processors and controllers \cite{Kelly2016}. In order to study tendencies about a population, data about its members have to be collected and protecting the privacy of the users is often an ethical and a legal responsibility for data controllers.

So how can citizen privacy be guaranteed while the data can still provide analytical value? Researchers propose Privacy Preserving Techniques (PPT) which allow the sharing of datasets with PPI while providing a certain degree of privacy or anonymity for individuals. However, the application of these PPT's often compromises the analytical value at the cost of privacy and vice-versa. RAPPOR \cite{Erlingsson2014} is a method that seeks to bridge this gap by applying Differential Privacy guarantees to collect data using an algorithm that relies on adjustable parameters that can provide varying levels of privacy guarantee. 

Anonymity and privacy may often be mistaken or interchanged. Pfitzmann \cite{Pfitzmann2001} defines anonymity as not disclosing the identity of a user reporting on actions, whereas privacy is the right to not disclosing whether an individual has performed a given action. As such, anonymity may be harmed given the existence of a known user in a dataset, while privacy is unharmed as long as that user's actions or answers remain unknown \cite{Harkiolakis2013}. Data collectors have traditionally applied PPT over released datasets through anonymisation and de-identification techniques. Unfortunately, these are often insufficient, as seen with the re-identification of AOL users from released, anonymised search queries \cite{Barbaro2006a}. Other examples include the identification of individuals via correlation of common demographic data such as zip codes, gender and date of birth \cite{Sweeney2000} or the de-anonymisation of sanitised data performed on the Netflix Prize data mining contest \cite{Narayanan2010a,Narayanan2006}.

This paper aims to observe the impact of such privacy parameter variations by applying the RAPPOR algorithm on three samples generated from a publicly available Android Malware analysis dataset \cite{Mirsky2016}, and thus evaluate the usability of differential privacy and to establish some configuration guidelines that satisfy both privacy guarantees and data controllers’ necessities. 

\section{Related Literature}\label{relatedLiterature}
Differential Privacy proposes a new, formal definition of privacy, and focuses on the study of population statistics while enforcing a strong deniability of individuals’ data through the addition of noise \cite{Dwork2006}. Being the only privacy guarantee that offers objective metrics about the level of confidentiality provided by a method \cite{Rossi2016}, the theoretical work started by Dwork et al \cite{Dwork2006} has increasingly been incorporated to new research and practical cases \cite{Wang2016a,Giakkoupis2015,Hu2015}. Its examples of integration include Apple, Microsoft and Google \cite{Greenberg2016,McSherry2009,Erlingsson2014a}. 

Differential Privacy is a definition of privacy whose ultimate paradigm states that no new knowledge of any individual should be learnt from accessing the data \cite{Dwork2013,Dwork2006a}. According to the differential privacy paradigm, the ability of an attacker to learn something about an individual should not be related to the presence or absence of the person in the dataset. This protection can thus be used encourage people to participate and give more honest answers, as the noise grants an objective level of maximum leakage, as well as a strong deniability against any conclusion that an adversary may extract from isolated records.

Although it is considered the gold standard in current privacy preservation research by authors like \cite{Hsu2014} or \cite{Huber2013}, some of the originators of core methods insist in saying that it is no panacea \cite{Dwork2011}, and that each case must be studied individually for good tuning of the parameters. As an example of its recognition amongst scholars, one of the papers where the method was initially presented \cite{Dwork2006} was given the Test-of-Time award \cite{Goldwasser2016}.

\section{Methodology}
The RAPPOR algorithm is conceived as a client-server architecture where the response of the clients will be automatically randomised before transferred to the server for the aggregation with the other responses \cite{Fanti2016}. This is a non-interactive process \cite{Dwork2011}, as the original data would not be recoverable. Using the RAPPOR algorithm, each response is encoded using Bloom filters \cite{Bloom1970} on which noise is added \cite{Erlingsson2014}. As part of the algorithm, the parameters can be adjusted to cater for various levels of privacy.

Noise is added to the Bloom filter following the surveying technique of randomised response \cite{Erlingsson2014}. This theory first proposed by \cite{Warner1965} and was designed to encourage participants in a survey to answer more honestly to sensitive questions: A coin was flipped to decide whether the answer to the question would be automatically \emph{yes} or the honest response, providing a strong deniability to such answer. Probability theory dictates that half of the population would have answered \emph{yes} regardless of the truth. Assuming that the other half of the answers were truthful and that the proportions in the population are represented accurately enough on the sample, the real proportion of \emph{no} would be the double of the proportion obtained from the survey.

After the Bloom filter is generated, two different randomisation phases are used to generate the final noisy response \cite{Erlingsson2014}. In this paper, we modify the following parameters:
\begin{itemize}
\item $f$ Probability of reporting lies.
\item $p$ Probability of reporting noise.
\item $q$ Probability of reporting a true answer.
\end{itemize}

\subsection{Data set considerations}
Taking the work of Erlingsson \cite{Erlingsson2014} as a reference, it can be extracted that, for $\epsilon=\ln(3)$, a dataset of 10,000 answers will be insufficient to extract useful information. A higher value of $\epsilon$ should guarantee an improved resemblance between the recovered information and the original data. Furthermore, the election of $\epsilon=\ln(3)$ is not justified in this original work, which seems to point that this level of protection is simply appropriate for testing. A question that naturally arises from this is: How different values of $\epsilon$ affect the accuracy of the data?

As Hsu et al. state \cite{Hsu2014}, a procedure for selecting values for $\epsilon$ that are: optimal; sufficient; and non-excessive is something that seems to be missing in the literature. This absence, at least partially, is due to the difficulty of estimating an optimal value for a single variable that must consider factors such as: the size and diversity of the dataset; the sensitivity of the information contained inside this; and the level of accuracy required for the reconstruction of the noisy response. In the design of such a system, the main disadvantage of the simplification of a scenario is the complexity of its optimisation.

Considering this situation, it appears to be evident that elaborating a general rule for predicting how the variation of $\epsilon$ will affect the privacy and fidelity levels of a specific dataset is something far from simple. An empirical demonstration of a progressive quality degradation of the recovered data as $\epsilon$ decreases could clarify this matter and help other researchers.

Erlingsson et al \cite{Erlingsson2014} also outline the case where there could be a loss of information due by an over-sized dataset and its consequent growth of noise after the RAPPOR process: Smaller, real values could be obfuscated by the noise due to their proportionally reduced presence, creating a situation where, literally, \emph{more is less}. In spite of being a logical conclusion, this affirmation is not supported by empirical data, which leads to wonder how and where that inflexion point could be found.

The paper thus aims to evaluate the effect of the following on an Android malware dataset:

\begin{itemize}
\item \emph{Investigate the value of $\epsilon$}. Decreasing the value of $\epsilon$ will offer noisier results, and these might not be even useful if the amount of data collected is not enough. How the variation of this parameter will be affected by the retrieved data, is thus key to the design of a procedure that offers reliable information to the data controller while protecting the privacy of users.
\item \emph{Investigate the size of the dataset}. For the same value of $\epsilon$, how will the results be affected on datasets with different sizes? Comparing results over different populations can help to estimate the minimum number of reports needed by different levels of privacy guarantee, as well as anticipating suitable configurations for practical cases. If the results obtained are comparable to those offered by Erlingsson et al \cite{Erlingsson2014}, this will also verify the procedures followed by them.
\end{itemize}

\subsection{Experimental Design}
The experiments proposed in this paper attempt to design different scenarios that will allow the study of the effects produced by the variation of privacy parameters. It will thus observe the differences between original data and data recovered from the RAPPOR reports, and also between data recovered from different scenarios, this is, the scenarios with the same number of malware sample reports but different values of $\epsilon$ and vice-versa.

RAPPOR requires a dataset with a considerable size and a wide variety of participants. Special interest is often put on finding data related to smartphones, as these devices contain a significant amount of potentially sensitive information about their users, and a considerable volume of data is collected from them on a regular basis \cite{Lord2012}. The chosen candidate is a smartphone dataset destined for security research \cite{Mirsky2016}. This dataset collected information about the devices of the participants, at the same time that a fake piece of malware was leaving traces. The election of this dataset is caused by the presence of process names as one of the pieces of data collected, establishing similarities between this experiment and the one performed by \cite{Erlingsson2014}. Process names are a suitable parameter for RAPPOR, as they are translated into a set of strings, some of which should have a very strong presence, such as system processes and common applications, while others should scarcely appear.

\subsubsection{Alteration of the Dataset}
The dataset used is partially available in public and fully by request \cite{Mirsky2016}. In this work, the partial public dataset was used, even in spite of the lack of variety of users, as there are more than enough pieces of data when sub-sampled. The field UUID, which represents the UNIX millisecond timestamp of the reports' collection, was used as a field by which the records were grouped by, using them as different RAPPOR users. As Listing \ref{countUniqueStrings_cli} shows, more than 300,000 users could be obtained using the UUID field (a considerable number compared to the 10,000 users from the second experiment made by Erlingsson et al \cite{Erlingsson2014}). This represents each user as a one-time collection of the status of the phone. With around 46 reports per user, compared to the 18 of Erlingsson et al \cite{Erlingsson2014}, this sample should be more than enough to obtain accurate results from RAPPOR. It could be argued that this adaptation would change the meaning of the data, but the experiment seeks to observe differences in the distributions of recovered data, and not to learn from it.

Aside from the two attributes mentioned before (UUID and process name), the rest of the data was obviated, as it was not necessary for the experiment, making it more simmilar to the original experiment. In the dataset, the column "PackageName" was used to name the running processes. While it is true that the column “ApplicationName" would have provided a more human-readable output, the former one was chosen as some applications contained unusual characters that could lead to code execution errors.

\begin{lstlisting}[caption=Count of unique strings in \cite{Mirsky2016},breaklines=true, label=countUniqueStrings_cli,]
>> awk 'BEGIN{FS=","}{print $2;}' Application.csv | sort | uniq | wc -l
>> 307051
\end{lstlisting}

\subsubsection{Election of Population Sizes}
In their first experiment, Erlingsson et al \cite{Erlingsson2014} used three different populations of 10,000; 100,000; and 1,000,000 responses to demonstrate the effect of the reports collected and the quality of the retrieved information. By using similar sizes, this experiment will allow the verification of their empirical work. Looking at the dataset that was chosen, even extracting one report from each user would produce a sample of 307,051 records, as seen in Listing \ref{countUniqueStrings_cli}. In order to produce smaller samples, a selection of users is chosen to obtain a reduced amount of records.

\subsubsection{Election of $\epsilon$ Values}
In differential privacy, the acceptable limits of the privacy guarantee parameter ($\epsilon$) range from 0.01 to 10 \cite{Hsu2014}. The value chosen by Erlingsson et al \cite{Erlingsson2014} is $\epsilon = \ln(3)$, which sits in the middle of this range, therefore being a parameter of medium privacy value. As the experiment aims to be illustrative for the changes suffered between the variations of the $\epsilon$ parameter, the high value will be extracted directly from the upper end (10) of the previously mentioned range in the literature, while the low value will be 0.1. Tests were run using 0.01, but no results at all were retrieved (0 detected occurrences of each candidate string). This fact is illustrative by itself on how low this value of $\epsilon$ may be for the chosen population sizes, but the same conclusions can be reached by using $\epsilon = 0.1$ (as it will be seen in the following sections), and which also offers more information to be discussed. Choosing 0.1 over 0.01 also offers an interesting point of view on the experiment, as the proportion between adjacent values would be 1:10.

\subsection{Evaluation of Results}
Some of the original work by Erlingsson et al \cite{Erlingsson2014} offer limited information about the results of their experiments, apart from the graphical comparisons between the original and the recovered distributions. The type of data that this paper will offer are discrete counts of appearances of each string with a simple method to evaluate the accuracy. Apart from the raw number of retrieved strings, each of the scenarios are evaluated according to the proportion of recovered strings that were detected with at least 80\% accuracy. The proportions are represented in comparison to the total of the strings detected, and not to the total of the strings present in the original data. In a real use case, the actual information of the population would not be available, and thus it is considered to be more illustrative representing the information in this manner, as it depicts how much of the retrieved information is indeed reliable. 

\section{Implementation}
\subsection{Code Analysis}
RAPPOR’s repository contains the core software for the server side of RAPPOR, which it is organised into a set of Python scripts that call R and C++ programs. Other sets of utilities like client libraries, scripts for server automation or web applications for easy simulations and data analysis are also present. A further inspection of the documentation (under the path doc/data-flow.md) revealed a highly detailed description of the data needed and the modules to run on them to obtain reports from original data, and a distribution from the reports. Unfortunately, a check of the pipeline described in the documentation (all scripted inside demo.sh) revealed that some modules had been changed since the documentation was created, and it is uncertain how these changes could affect the original results.

\subsection{Maximum Size of the Bloom Filters}
Both the Python and the C++ clients offered in the repository possess a limit of 32 bits for the size of the Bloom filter used to generate the reports. In spite of this, according to the experiments shown on the original paper, the sizes of the Bloom filters used in them is considerably larger. 

In \cite{Erlingsson2014}, the first experiment uses Basic One Time RAPPOR, which requires an individual bit for every candidate string, and considering that 200 candidates were employed and that the size of the filter has to be a number such as $x = 2^n$, the minimum size of this filter had to be 256 bits. The second, third and fourth experiments used Bloom filters of 128 bits. Thus, it appears that the code used by Erlingsson et al \cite{Erlingsson2014} may differ from the one offered in the GitHub repository. Consequently, reproducing the exact scenario with the code available is not possible. However, according to the experiments made by \cite{Erlingsson2014}, the size of the Bloom filter should not affect the retrievability of the data, as it does not alter the value of $\epsilon$. A 32-bit Bloom filter with two hashes (used in this work) offers $32^2 = 1024$ different combinations, more than enough compared to the 154 candidate strings present in the experimental datasets.

\subsection{Code Creation}
A script \cite{Acuna2018} was created to automate the process of applying RAPPOR to a dataset and then retrieve meaningful data from its outcome. This script essentially takes the necessary components of the original demo and automated the process to speed up the process.

\subsection{Dataset Creation and sub-sampling}
The data needed for feeding the pipeline consists on a comma separated file with a header and two columns. The first column is a unique identifier for every client, and the second contains the value to encode. Using the previously described dataset (named Applications.csv) \cite{Mirsky2016}, the AWK command on Listing \ref{awkDatasetSubset} was performed to extract the data (\textit{dataset.csv}) required for further sub-sampled datasets. The CUT command on the same listing was used to remove blank lines and extract a \textit{unique.txt} file with unique strings within the extracted dataset.

\begin{lstlisting}[caption=Data extraction from Application.csv,breaklines=true,label=awkDatasetSubset]
>> awk 'BEGIN{FS=","}{print $2","$5:}' Applications.csv > dataset.csv
>> cut -d ',' -f 5 Applications.csv | sort | uniq > uniques.txt
\end{lstlisting}

From this list, the before mentioned blank space and the header “PackageName” had to be removed via manual inspection. A script that creates subsamples from aleatory selections of reports was elaborated to create the datasets for the experiment \cite{Acuna2018}. This script enables to create such sub-samples choosing the number of users and reports per user to select. In this way, the populations of 10,000 and 100,000 pieces of data were created by selecting just one report from users until completing the list, while the population of 1,200,000 pieces of data was constructed selecting four reports from 300,000 users.

\section{Experimental Execution}
The values of $\epsilon$ decided on the methodology were approximated using the $\epsilon$ calculator from \cite{Acuna2018}. As the number of hashes affects the recoverability conditions \cite{Erlingsson2014}, this value was left unchanged and set to two hashes, the same number used on the three of the experiments performed by Erlingsson et al \cite{Erlingsson2014}. Table \ref{epsilonCalc} show the configurations chosen for the experiment. Each of these sets of values was collected in three different files with the same structure. The files were named  \textit{params\_01.csv}, \textit{params\_1.csv} and \textit{params\_10.csv}.

\begin{table}[htbp]
\centering
\caption{Values chosen to obtain different $\epsilon$ for the experiment}
\label{epsilonCalc}
\begin{tabular}{@{}lllll@{}}
\toprule
$\epsilon$       & f    & h & p    & q    \\ \midrule
0.1     & 0.75 & 2 & 0.5  & 0.55 \\
1.0743  & 0.50 & 2 & 0.5  & 0.75 \\
10.0184 & 0.01 & 2 & 0.05 & 0.90 \\ \bottomrule
\end{tabular}
\end{table}

\subsection{Report Generation}
Using the parameter files, different series of reports were generated using the \textit{pipeline.sh} script from \cite{Acuna2018}. This script runs the pipeline for the three different datasets and the three different parameter groups, creating a total of nine different scenarios. After the generation of the scenarios, each set of files (params.csv, counts.csv and map.csv) was uploaded into the original analysis web application from \cite{Erlingsson2014a}. The application offered a summary file with estimation data and a results file with the estimation of the appearance of the different candidate strings.

The file \textit{results.csv}, together with the original dataset for that concrete experiment, was used to generate the final comparison file through the script depicted using a summary generator \cite{Acuna2018}. The resultant file compares the counts of candidate strings in the original distribution versus the estimation. Therefore, this file is suitable for being used on plotting software to obtain the final graphics.

\section{Evaluation}
To establish an objective and precise evaluation methodology, the results were compared by two criteria: Cases with different population size but identical $\epsilon$ values, and cases with different $\epsilon$ value on the same population.

\subsubsection{Data Presentation}
The information collected from the experiment is shown in Figures \ref{exp1_e1_p10k_fig}-\ref{PPV_exp2_fig}. These figures will be referenced throughout the section while the obtained results are evaluated. In order to summarise the data extracted from the experiment, Table \ref{PPV_exp1_tbl} depicts the number of strings detected by the analysis after these had been encoded in RAPPOR reports.

\begin{table}[!t]
\centering
\caption{Raw results of high, medium, and low privacy guarantees}
\label{PPV_exp1_tbl}
\resizebox{0.5\textwidth}{!}{%
\begin{tabular}{@{}rcccr@{}}
\toprule
Population & $\epsilon$ & True strings & RAPPOR strings & 80\% accuracy \\ \midrule
10k & 0.1 & 121 & 0 & 0 \\
10k & 1.0 & 121 & 7 & 1 \\
10k & 10.0 & 121 & 39 & 23 \\ \midrule
100k & 0.1 & 140 & 1 & 0 \\
100k & 1.0 & 140 & 14 & 7 \\
100k & 10.0 & 140 & 60 & 40 \\ \midrule
1,200k & 0.1 & 143 & 2 & 0 \\
1,200k & 1.0 & 143 & 47 & 24 \\
1,200k & 10.0 & 143 & 77 & 60 \\ \bottomrule
\end{tabular}
}
\end{table}

\subsubsection{Data Analysis}
In order to illustrate more objectively the results obtained, a statistical collector script \cite{Acuna2018} calculates the number of detected strings that approximated the real count with a 20\% maximum margin of error, as defined in Table \ref{PPV_exp1_tbl}. This value is chosen to discard values too distant to the original one.

\subsection{Varying $\epsilon$ Values}
The value of $\epsilon$ determines entirely the level of protection provided by differential privacy. According to McSherry \cite{McSherry2010}, an $\epsilon\le0.1$ should be considered as a strong guarantee of privacy, while an $\epsilon\ge10$ should be considered as a weak guarantee. In this section, these ``strong'' and ``weak'' levels will be tested with the previously generated datasets, to evaluate the quality of the recovered data that RAPPOR can offer in these situations. An intermediate level of $\epsilon=1$, equidistant to the other two values by a proportion of 1:10, is also included to show the transition and appreciate better the changes that the recovered data suffer as the value of $\epsilon$ varies.

\subsubsection*{$\epsilon=0.1$}
Table \ref{PPV_exp1_tbl} shows how none of the recovered pieces of data from the smaller $\epsilon$ value could retrieve a single string with less than a 20\% margin of error. Even being ten times higher than the smallest value found in literature \cite{Sarwate2010}, this restriction appears to be too constraining to be useful in, at least, datasets of the sizes used for this experiment. This also confirms the conclusions of \cite{McSherry2010}, saying that such a low value for $\epsilon$ produces considerably noisy responses, prone to offer a very high false positive rate. In the case of Sarwate et al \cite{Sarwate2010}, it could be justified the use of such a small value, as their project focuses on machine learning, a field where processing substantially larger datasets are rather common. Nevertheless, without further testing, it is not possible to determine whether the differential privacy mechanism concretely implemented by RAPPOR would successfully retrieve useful information in such conditions.

\subsubsection*{$\epsilon=1.0$}
\begin{figure}[!t]
\centering{\includegraphics[width=\columnwidth]{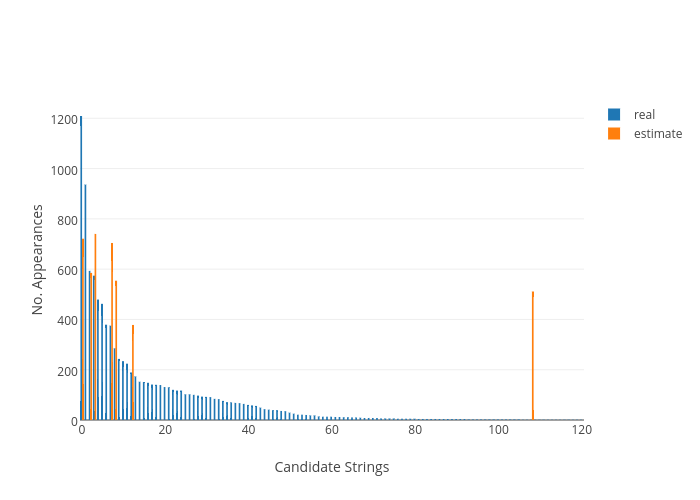}}
\caption{Noise on 10,000 responses using $\epsilon = 1$}
\label{exp1_e1_p10k_fig}
\end{figure}

\begin{figure}[!t]
\centering{\includegraphics[width=\columnwidth]{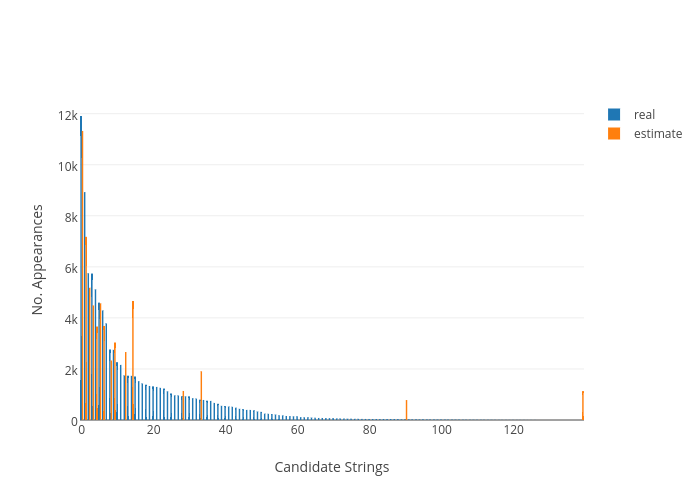}}
\caption{Noise on 100,000 responses using $\epsilon = 1$}
\label{exp1_e1_p100k_fig}
\end{figure}

\begin{figure}[!t]
\centering{\includegraphics[width=\columnwidth]{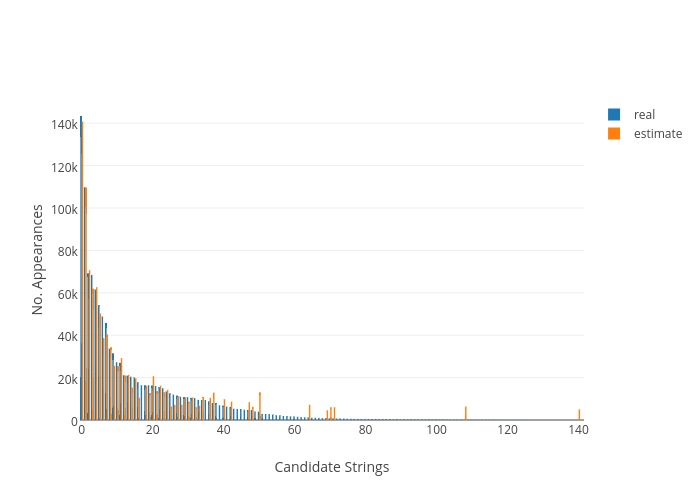}}
\caption{Noise on 1,200,000 responses using $\epsilon = 1$}
\label{exp1_e1_p1000k_fig}
\end{figure}

Replicating the same privacy parameters used in the experiments of Erlingsson et al \cite{Erlingsson2014}, Figures \ref{exp1_e1_p10k_fig}-\ref{exp1_e1_p1000k_fig} show how significant the size of the population is when choosing privacy values on RAPPOR. On the smallest population (Figure \ref{exp1_e1_p10k_fig}), only some of the most popular strings are detected, with a clear false positive at the tail of the distribution. Only \sfrac{1}{7} of the detected values are accurately approximated, as given in Table \ref{PPV_exp1_tbl}.

In the medium population of 100,000, shown in Figure \ref{exp1_e1_p100k_fig}, the most popular strings are properly retrieved to a certain level of acceptance, with 50\% of the detected values within the range of 20\% error. The tail still shows some false positive results, but they are smaller in proportion and more evenly distributed over the candidate strings. Despite the fact that the head of the estimated distribution seems to start shaping closer to the real data, only a few strings come close enough to the original, being difficult to accurately determine even the heavy hitters.

Figure \ref{exp1_e1_p1000k_fig} shows how retrievability increase significantly given greater amounts of data. In spite of this, with 10 times more data than the previous sample, the growth of the proportion of strings detected within 20\% becomes more steady. This situation, inverted from the one described in the transition from 10,000 to 100,000 users, could imply that a higher increase in the number of reports is largely beneficial to detecting new strings, but not as helpful in increasing the accuracy of recovered data.

\subsubsection*{$\epsilon=10$}
With the weakest privacy guarantee, on the smallest dataset, 32\% of the original strings are recovered and 59\% of these within 20\% error margin. In comparison, on the biggest dataset 54\% of the original strings are recovered, 59\% of these within the 20\% error margin of the true number of appearances. Thus, the number of accurate strings recovered may not increase drastically between the two data sizes, but there are significantly more strings detected on the larger population. As it was expected from what was learnt from literature, the original and the estimated distributions match almost perfectly. 

The amount of detected strings has a small growth from the previous population size, compared to the growth of strings detected with an 80\% of precision, which could indicate that the strings with smaller appearances cannot be detected even with disproportionately sized datasets, while increasing the number of reports collected contributes significantly to the accuracy of the strings detected.

In the 100,000 population and with $\epsilon = 1$, the recovered distribution already adopts very closely the shape of the original dataset, as Figure \ref{exp1_e1_p100k_fig} shows. 66.7\% of the recovered strings are inside the 20\% margin of error.

\subsection{Same Population Sizes}
The samples exposed earlier will be now grouped by population size, with the intention of evaluating the effects of altering $\epsilon$ on the same original distribution, and judge how this affects the retrievability of the data.

\subsubsection*{Population of 10,000}
In the transition from $\epsilon = 0.1$ to $\epsilon = 1$, only 14.3\% of the strings appear to be detected with precision. It is interesting to notice that, according to Figure \ref{exp1_e1_p10k_fig}, the most important strings in the original distribution would not be present in these groups, as their estimated values are considerably distant from the original. This could be due to the concentration of these values on a reduced number of cohorts, making the detection more difficult with this level of privacy guarantee. Passing from $\epsilon = 1$ to $\epsilon = 10$, the proportion of detected strings grows acceptably, while 59\% of these are within the 20\% margin of error and thus providing a highly accurate detection rate.

\subsubsection*{Population of 100,000}
A high growth of the 20\% limit is shown on the first transition from 0.1 to 1 $\epsilon$ value, going from 0\% to 50\%. Figure \ref{exp1_e1_p100k_fig} shows how the most accurate values concentrate around the head of the distribution, while the tail remains barely detected. Changing $\epsilon$ to 10 causes the estimation to grow considerably closer to the original data, with a more modest growth in the 20\% error limit.

\subsubsection*{Population of 1,200,000}
Even the largest of the populations show no successful results when being processed through the more restricting privacy parameter. Although the detection of the biggest string may be seen, not even this value is approximated within at least 80\% accuracy. The transition to $\epsilon = 1$ recreates the head of the distribution rather accurately (Figure \ref{exp1_e1_p1000k_fig}), with 27.6\% of the strings accurately detected. The least demanding privacy restriction provides 77.9\% of the strings detected with less than a 20\% error.

\subsection{Further details of high privacy guarantees}
\begin{figure}[!t]
\centering{\includegraphics[width=0.5\textwidth]{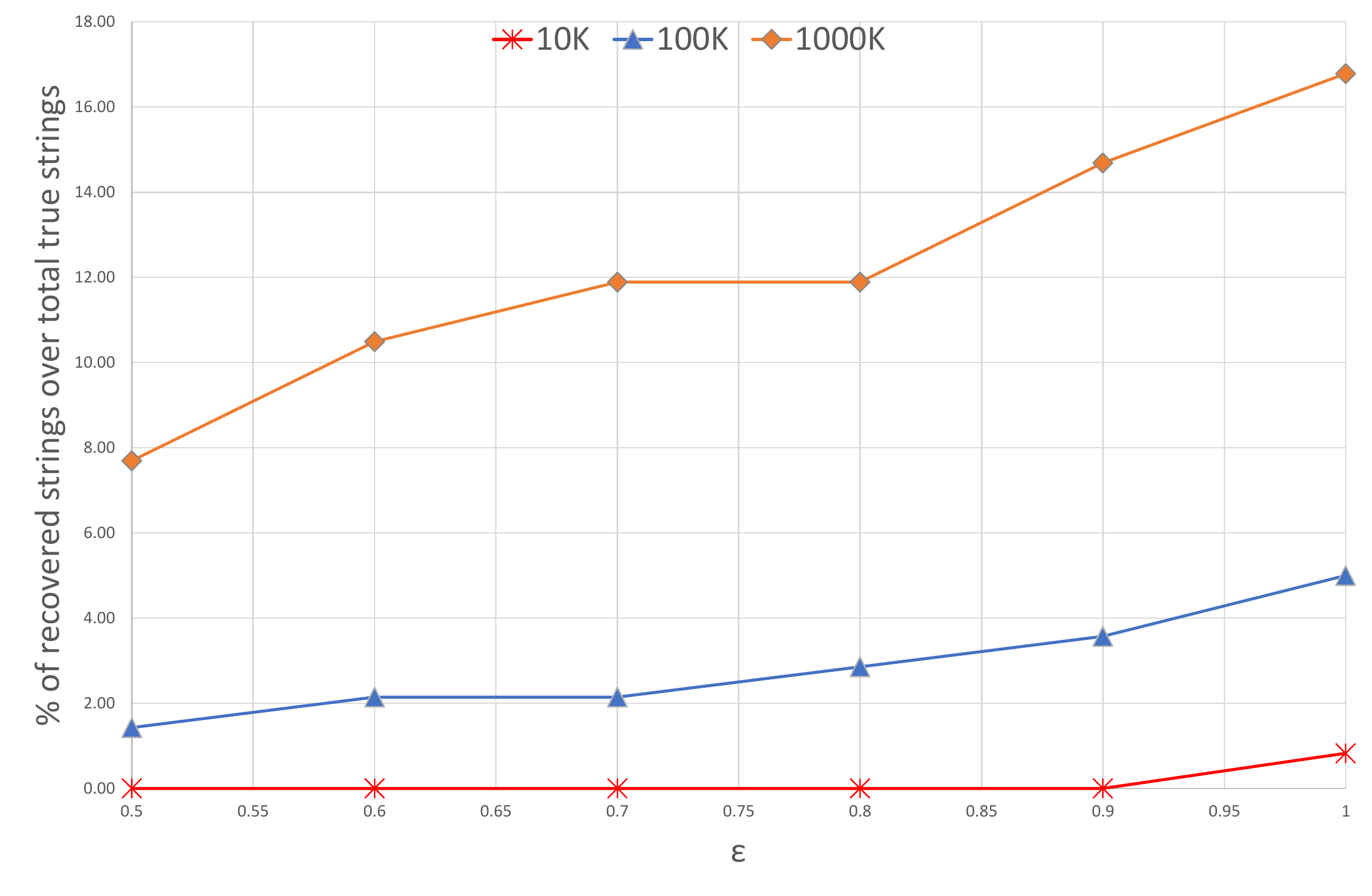}}
\caption{Privacy Parameter Variation comparison between population sizes}
\label{PPV_exp2_fig}
\end{figure}

As seen in the paper by Erlingson et al \cite{Erlingsson2014}, limited privacy guarantees were tested against three different population sizes. Additionally, privacy guarantees were selected based on popular values in literature and did not present detailed results on varying the privacy parameters. This section outlines the results of applying RAPPOR on the three datasets with varied privacy guarantees between $\epsilon = 0.1 - 1$. Similar to the results previously exposed, which are comparable and verifiable to the results in \cite{Erlingsson2014}, this evaluation will be based on the raw results in Table \ref{PPV_exp2_tbl}. Looking at Table \ref{PPV_exp2_tbl}, six results are shown for each population size, and further visualisation of these can be seen in Figure \ref{PPV_exp2_fig}. Figure \ref{PPV_exp2_fig} outlines a scatter-plot of the strings retrieved with 80\% accuracy observed for each $\epsilon$ variation, normalised over the true number of strings in each population for each dataset. 

\subsubsection{Different impact on different population sizes}
The increased privacy guarantees affect the smallest dataset of 10,000 reports significantly, with no strings accurately detected with strict privacy guarantees. This is, however, expected, as there was just one string discovered in the previous experiment (Table \ref{PPV_exp1_tbl}), and stricter privacy guarantees would generate greater levels of noise. The RAPPOR strings detected increase and decrease through the varied privacy parameters, but never approach the 80\% accuracy barrier at any point for the small dataset, except from $\epsilon = 1$. When observing the results of the 100,000 population, interesting conclusions can be extracted, as the impact of decreasing privacy guarantees affects the recovery greatly at first, but then its growth slowly decays when arriving to the more relaxed privacy guarantee values. It may appear that the analytical value of using $\epsilon = 0.6$ over $\epsilon = 0.7$ remains the same even considering that the privacy guarantee decreases. Nevertheless, as it was previously mentioned, differential privacy and the RAPPOR algorithm performs better on larger populations, with these inconsistencies being due to the randomisation factor of the process. When observing the 1,200,000 response data, a similar behaviour can be seen when observing the impact of $\epsilon = 0.6$ and $\epsilon = 0.7$. However, the overall accuracy is severely affected in comparison to the smaller dataset. Logically, the smaller dataset is already performing poorly and thus cannot be affected further, in contrast to the larger dataset, which has over 15\% strings detected using $\epsilon = 1.0$ and down to 0.08\% when increasing the privacy guarantees.

\begin{table}[!t]
\centering
\caption{High Privacy Parameter Variation results}
\label{PPV_exp2_tbl}
\resizebox{0.5\textwidth}{!}{%
\begin{tabular}{@{}rcccr@{}}
\toprule
Population & $\epsilon$ & True strings & RAPPOR strings & 80\% accuracy \\ \midrule
10k & 0.5 & 121 & 1 & 0 \\
10k & 0.6 & 121 & 2 & 0 \\
10k & 0.7 & 121 & 0 & 0 \\
10k & 0.8 & 121 & 3 & 0 \\
10k & 0.9 & 121 & 1 & 0 \\
10k & 1.0 & 121 & 7 & 1 \\ \midrule
100k & 0.5 & 140 & 8 & 2 \\
100k & 0.6 & 140 & 11 & 3 \\
100k & 0.7 & 140 & 12 & 3 \\
100k & 0.8 & 140 & 16 & 4 \\
100k & 0.9 & 140 & 15 & 5 \\
100k & 1.0 & 140 & 14 & 7 \\ \midrule
1,200k & 0.5 & 143 & 22 & 11 \\
1,200k & 0.6 & 143 & 26 & 15 \\
1,200k & 0.7 & 143 & 28 & 17 \\
1,200k & 0.8 & 143 & 33 & 17 \\
1,200k & 0.9 & 143 & 35 & 21 \\
1,200k & 1.0 & 143 & 47 & 24 \\ \bottomrule
\end{tabular}
}
\end{table}

\section{Conclusions and future work}
This paper defined an experiment that created different scenarios for altering the RAPPOR input values, and verified that the results obtained by Erlingsson \cite{Erlingsson2014} in their experiments are repeatable, ensuring the quality of their work. However, utmost care must be taken when deciding on the privacy parameters of RAPPOR. It was discovered that $\epsilon = 0.1$ was a privacy guarantee too high to be met even in the most optimistic of the scenarios, so it remains uncertain whether this value would be usable for bigger populations. The 10,000 responses population only offered reliable results when the privacy guarantee was lowered to almost non-existent levels ($\epsilon = 10$). It was appreciated that, after a certain threshold of population size, data started to be meaningful, observing great increases of accuracy from $\epsilon = 0.7$ and higher for both the 100,000- and 1,200,000-sized datasets. After that initial growth, the rise of acceptable detected strings slowed, starting to increase the precision of the already discovered data instead. This lead to the conclusion that, from the aforementioned threshold onwards, collecting more reports would increase the quality of the retrieved data, but not the raw number of detected strigs.

One area of development is the difficulty of generating Bloom filters larger than 32 bits. Even though the results from Erlingsson et al \cite{Erlingsson2014} have been successfully verified, this difficulty prevents the replication of the exact conditions of their empirical work. Altering the code to make this feature available would significantly improve the work done in this paper. Also, this would help to create a more versatile tool, adaptable to different situations. Additional tests over different Bloom filter settings would also verify that the variation of the size of the filters does not affect the recoverability of the reports, as \cite{Erlingsson2014} assure. 

The smallest of the chosen values of $\epsilon$ showed that is was impractical for populations of up to 1,200,000 reports. A possible additional check would be to determine if such a constraining privacy guarantee can be used with larger populations, or if, on the contrary, this value makes RAPPOR generate reports that are too noisy to be of any use.

Choosing three different sizes, the sub-sampled datasets were created by randomly selecting one report per user for the two smaller populations, and four different reports per user for the largest one. Testing populations of the same size with a different number of users and responses per user (e.g., one response for 1000 users, two responses for 500 users, four responses for 250 users\ldots) could help to further understand the influence of the characteristics of the population in the recoverability of the reports. 


\bibliographystyle{IEEEtran} 
\bibliography{bibliography.bib}
\end{document}